\documentclass{pasj00}
\SetRunningHead{T. Nagata et al.}{X-Ray Nova XTE J1720$-$318}
\Received{2003/07/23}
\Accepted{2003/10/22}
\Published{}
\begin{document}

\title{
The Infrared Counterpart of the X-Ray Nova XTE J1720$-$318
}
\author{Tetsuya \textsc{Nagata}, Daisuke \textsc{Kato},
Daisuke \textsc{Baba}, Shogo \textsc{Nishiyama}, 
Takahiro \textsc{Nagayama}, \\
Chie \textsc{Nagashima}, Mikio \textsc{Kurita}, and Shuji \textsc{Sato}}
\affil{Department of Physics, Nagoya University, Chikusa, Nagoya 464-8602}
\email{nagata@z.phys.nagoya-u.ac.jp}
\author{Taichi \textsc{Kato} and Makoto \textsc{Uemura}}
\affil{Department of Astrophysics, Kyoto University, Sakyo, Kyoto 606-8502}
\author{Hitoshi \textsc{Yamaoka}}
\affil{Department of Physics, Kyushu University, Fukuoka 810-8560}
\author{Berto \textsc{Monard}}
\affil{Bronberg Observatory, 
 PO Box 11426, Tiegerpoort 0056, South Africa}
\author{Yoshifusa \textsc{Ita} and Noriyuki \textsc{Matsunaga}}
\affil{Institute of Astronomy, School of Science, 
 The University of Tokyo, Mitaka, Tokyo 181-0015}
\author{Yasushi \textsc{Nakajima} and Motohide \textsc{Tamura}}
\affil{National Astronomical Observatory, Mitaka, Tokyo 181-8858}
\author{Hidehiko \textsc{Nakaya}}
\affil{Subaru Telescope, National Astronomical Observatory of Japan, 
Hilo, HI 96720, USA}
\and 
\author{Koji \textsc{Sugitani}}
\affil{Institute of Natural Sciences, Nagoya City University, 
Mizuho, Nagoya 467-8501}
\maketitle

\begin{abstract}
We report on the discovery of an infrared counterpart 
to the X-ray transient XTE J1720$-$318 on 2003 January 18, 
nine days after an X-ray outburst, 
and the infrared light curve during the first 130 days 
after the outburst.  
The infrared light curve shows a decline of $\sim$1.2 mag 
from the peak magnitude of $K_{\mathrm s} \sim15.3$ over the observation period, 
and a secondary maximum, 
about 40 days after the outburst.  
Another small increase in the flux was also recorded 
about 20 days after the outburst.  
These increases were also detected in the X-ray light curve.
The $J H K_{\mathrm s}$ colors are consistent with 
an X-ray irradiated accretion disk suffering 
an extinction of $A_V \sim 8$, 
which is also inferred from its X-ray spectrum and 
the extinction map constructed from far-infrared dust 
emission of this line of sight.  
These $J, H,$ and $K_{\mathrm s}$ observations 
demonstrate that useful data can be obtained 
even for such an object, 
which suffers heavy optical extinction, 
possibly located beyond the Galactic center.  

\end{abstract}

\section{Introduction}
X-ray novae (XNe), also called soft X-ray transients, are
semi-detached binary systems which experience luminous outbursts
observed in all wavelengths (Tanaka, Shibazaki 1996).  Most of them
contain a stellar-mass black hole as their primary (Liu et al. 2001).
Their brightness variations generally originate from an accretion
disk, and hence they can provide an ideal laboratory to study the
physics of accretion onto a black hole and the black hole, itself.  

X-ray light curves of typical XNe are characterized by a Fast Rise of a
few days and an Exponential Decay (FRED) with an $e$-folding time of 
$\sim 40$ d (Chen et al. 1997).  During an exponential decay, XNe
often exhibit a secondary maximum, also called a reflare observed 
30--50 days after the peak (Tanaka, Shibazaki 1996).  The accretion-disk
instability model, which was originally developed for dwarf nova
outbursts, has been applied to explain this type of outburst behavior
(Mineshige, Wheeler 1989; Ichikawa et al. 1994; King, Ritter 1998;
Truss et al. 2002), while the nature of reflares and various types of
outburst light curves are still open issues (\cite{chen93}; 
\cite{augu93}; \cite{mccl03}).  Simultaneous
observations of X-ray and optical--IR emissions are important because
they can provide crucial clues concerning the nature of the emission source, 
the outburst mechanism, and the propagation of hot regions in the disk
(e.g., Hameury et al. 1997; Uemura et al. 2000; Wu et al. 2002).

A new XN, XTE J1720$-$318 was discovered with the All-Sky Monitor (ASM)
onboard the Rossi X-Ray Timing Explorer (RXTE) at $130\pm20$ mCrab
(2--12 keV) on 2003 January 9 (Remillard et al. 2003).  Shortly after the
X-ray discovery, a radio counterpart was detected at
\timeform{17h19m59s.062} ($\pm$\timeform{0s.087}), 
\timeform{-31D44'59''.7} ($\pm$\timeform{1''.10})
(Rupen et al. 2003).  No new optical source brighter than $R_{\rm
c}\sim 18.0\;{\rm mag}$ was detected in CCD images taken on 2003
January 16 at the radio position.
\footnote{$\langle$http://vsnet.kusastro.kyoto-u.ac.jp/vsnet/Mail/vsnet-campaign-xray/msg00179.html$\rangle$, 
$\langle$http://vsnet.kusastro.kyoto-u.ac.jp/vsnet/Mail/vsnet-campaign-xray/msg00182.html$\rangle$.}
After reaching the X-ray maximum of $\sim 400$ mCrab (2--12keV) around
January 10, it started fading.  Its X-ray spectrum was first moderately
hard during January 9--10, and then became relatively soft (Remillard et
al. 2003).   After a softening of the spectrum, the high-frequency
variability was reported to be low, which strongly indicates that the
object entered a high/soft state often observed in the black hole XNe
(Markwardt, Swank 2003; Markwadt
\footnote{ATEL\#115, $\langle$http://atel.caltech.edu/$\rangle$.}).  
However, our knowledge of this source is still limited.  

We discovered an infrared counterpart of XTE J1720$-$318 on 
2003 January 18 (Kato et al. 2003).  
Here, we report on the time evolution of the flux and colors of this IR 
source, and the correlation between the X-ray and IR light curves.  

\section{Observations}
We obtained photometry at 
the near-infrared wavelengths $J$ (1.25 $\mu$m), $H$ 
(1.63 $\mu$m), and $K_{\mathrm s}$ (2.14 $\mu$m) using 
the near-infrared camera SIRIUS 
(Simultaneous three-color Infra Red Imager
for Unbiased Survey) on the IRSF (Infra Red Survey 
Facility) 1.4 m telescope of Nagoya University at 
Sutherland, South African Astronomical Observatory.  
Thirteen observations were made between 2003 January 18 
and 2003 May 21 UT.  

SIRIUS is equipped with three 1024$\times$1024 pixel HgCdTe arrays.  
Dichroic mirrors enable simultaneous observations in the three bands; 
thus, even if the source shows a rapid fluctuation in flux, 
its color information is correctly recorded.  
Details of the camera are given by 
\citet{naga99} and \citet{naga03}.
The image scale of the array is \timeform{0''.45} pixel$^{-1}$, 
giving a field of view of \timeform{7'.7}$\times$\timeform{7'.7}.  

Each night we repeated a set of observations several times 
with 10 different dithered positions, 
which resulted in a total integration time of 1200 s.  
The observations were made at various air masses, 
with some as large as 2.9 in the first few night observations.  
Typical seeing conditions were \timeform{1''.3} (FWHM) 
in the $J$ band.  
We observed the standard star 9172 in the 
faint near-infrared standard star catalog of Persson 
et al.(1998) for photometric calibration on 2003 February 14.  
The preceding and subsequent photometry of the infrared counterpart 
of XTE J1720$-$318 was obtained relative to 
several thousand stars around XTE J1720$-$318 
in the \timeform{7'.7} field.  
The IRSF/SIRIUS instrumental magnitudes are thus based on 
the assumption that Persson 9172 is $J=12.48$, $H=12.12$, $K_{\mathrm s}=12.03$; 
they can be transformed 
to the CIT system with the color equations,
\footnote{$\langle$http://optik2.mtk.nao.ac.jp/$\sim$yas/color/IRSFcolor-e.html$\rangle$.}
but we have not made the transformation here.

We applied the standard procedures of near-infrared array image 
reduction, 
including dark-current subtraction, sky subtraction, 
and flat-fielding, using the IRAF (Imaging Reduction and 
Analysis Facility)
\footnote{IRAF is distributed by the National Optical Astronomy Observatory, 
which is operated by the Association of Universities for Research in 
Astronomy, Inc., under cooperative agreement with the National Science. }
software package.
Identification and photometry of point sources were performed 
by using the DAOPHOT package in IRAF.  
We generally used a radius of $\sim$3 pixel (\timeform{1''.35}) 
in the PHOT procedure PSF fitting 
so that the nearby stars did not affect the photometry.  
In particular, a star \timeform{1''} in the west clearly detected in 
the VLT observation (see below)
seemed to be $K_{\mathrm s}\sim18$ or fainter and 
to have less than a 1/5 contribution to the flux, 
even when the object was dim in May. 

\begin{figure}
   \begin{center}
      \FigureFile(80mm,50mm){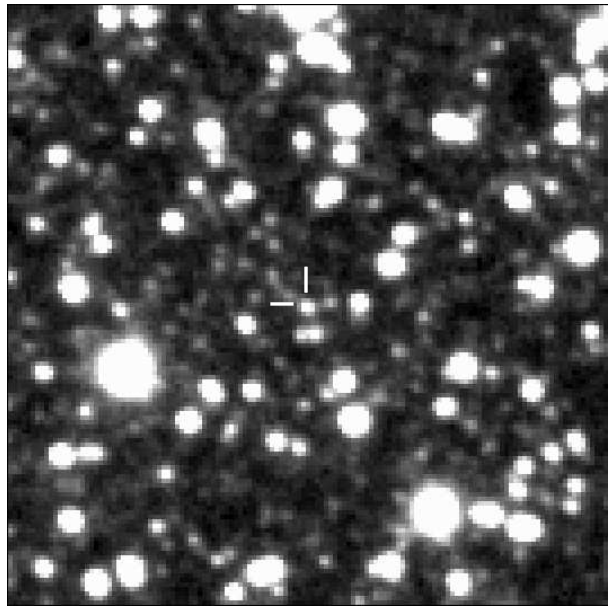}
   \end{center}
   \caption{$K_{\mathrm s}$ image of XTE J1720$-$318 on 2003 January 19.  
   The central \timeform{1'} part of 
   the IRSF/SIRIUS \timeform{7'.7} field is shown.}\label{figimage}
\end{figure}

\section{Results and Discussion}
The infrared counterpart was detected (figure \ref{figimage}) at 
\timeform{17h19m59s.00},
\timeform{-31D45'01''.2} 
(equinox 2000.0; determined from $\sim$150 2MASS reference
stars; rms error \timeform{0".2}), 
close to the radio counterpart (Rupen et al. 2003).  
O'Brien et al.
\footnote{ATEL\#117, ibid.} 
then confirmed it by VLT infrared observations, 
%IR 17:19:58.994, Dec = -31:45:01.25
and also reported an improved radio position of 
\timeform{17h19m58s.985},
\timeform{-31D45'01''.109} 
with a total positional uncertainty of \timeform{0''.25};
thus, the radio and infrared counterpart positions agree.  
%It has a several fainter neaby (whose separation is < \timeform{2''}) sources, 
It is invisible on public 2MASS images, 
and an examination of five $B$ and $R$ historic plates provided by the USNO B1.0
DSS Image and Catalogue Archive (POSSI, SRCJ, ESOR, and AAOR surveys)
showed nothing at this position down to $R \sim$20.0; $B \sim$21.0.
\footnote{$\langle$http://vsnet.kusastro.kyoto-u.ac.jp/vsnet/Mail/vsnet-campaign-xray/msg00182.html$\rangle$.}
The optical--IR magnitudes of the quiescent state 
and the amplitude of the infrared outburst are thus undetermined.  

\begin{figure}
   \begin{center}
      \FigureFile(88mm,55mm){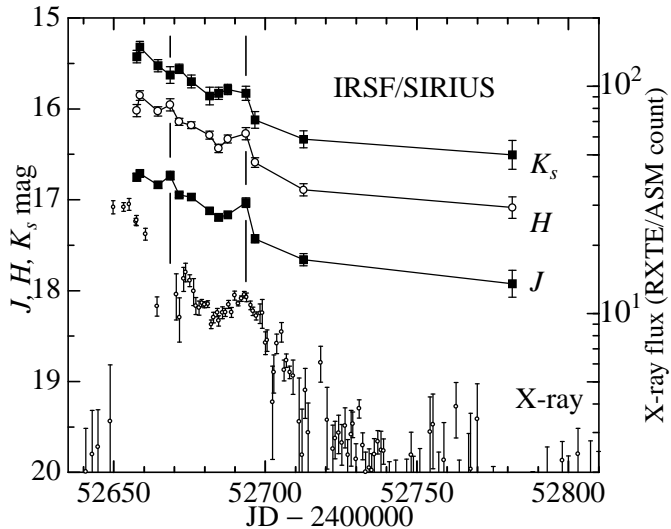}
   \end{center}
   \caption{Light curve of XTE J1720$-$318. 
   The X-ray data (2--12keV) are 
   from the RXTE/ASM public archive on the web 
   $\langle$http://xte.mit.edu$\rangle$.  
   The vertical lines indicate January 29 and February 23,
   when the infrared flux showed secondary maxima. 
   }\label{figlightc}
\end{figure}

\begin{table*}
\begin{center}
\caption{Near-infrared magnitudes of XTE J1720$-$318.}
 \begin{tabular}{ccccc}
% \begin{tabular*}{85mm}{@{\extracolsep{\fill}}ccccc}
  \hline\hline
     JD $-$   & Date UT    & $J$   & $H$   & $K_{\mathrm s}$   \\
     2400000  & (2003)     & (mag) & (mag) & (mag)   \\
  \hline
52657.6 & Jan 18 02:26--02:59 & 16.75{\scriptsize $\pm$0.04} & 16.02{\scriptsize $\pm$0.07} & 15.43{\scriptsize $\pm$0.07} \\
52658.6 & Jan 19 02:23--02:56 & 16.71{\scriptsize $\pm$0.03} & 15.85{\scriptsize $\pm$0.05} & 15.32{\scriptsize $\pm$0.06} \\
52664.6 & Jan 25 02:34--03:21 & 16.84{\scriptsize $\pm$0.04} & 16.03{\scriptsize $\pm$0.05} & 15.53{\scriptsize $\pm$0.07} \\
52668.6 & Jan 29 02:30--03:17 & 16.73{\scriptsize $\pm$0.05} & 15.96{\scriptsize $\pm$0.07} & 15.63{\scriptsize $\pm$0.09} \\
52671.6 & Feb 01 02:33--03:21 & 16.95{\scriptsize $\pm$0.03} & 16.14{\scriptsize $\pm$0.04} & 15.56{\scriptsize $\pm$0.05} \\
52675.6 & Feb 05 02:22--03:12 & 16.97{\scriptsize $\pm$0.03} & 16.18{\scriptsize $\pm$0.04} & 15.70{\scriptsize $\pm$0.07} \\
52681.6 & Feb 11 02:29--03:19 & 17.12{\scriptsize $\pm$0.04} & 16.29{\scriptsize $\pm$0.04} & 15.86{\scriptsize $\pm$0.10} \\
52684.6 & Feb 14 02:43--03:33 & 17.19{\scriptsize $\pm$0.03} & 16.44{\scriptsize $\pm$0.05} & 15.83{\scriptsize $\pm$0.07} \\
52687.6 & Feb 17 02:31--03:15 & 17.17{\scriptsize $\pm$0.03} & 16.33{\scriptsize $\pm$0.04} & 15.79{\scriptsize $\pm$0.06} \\
52693.6 & Feb 23 01:25--02:06 & 17.03{\scriptsize $\pm$0.05} & 16.27{\scriptsize $\pm$0.07} & 15.83{\scriptsize $\pm$0.08} \\
52696.6 & Feb 26 01:30--02:16 & 17.43{\scriptsize $\pm$0.04} & 16.59{\scriptsize $\pm$0.05} & 16.12{\scriptsize $\pm$0.09} \\
52712.7 & Mar 14 03:22--04:05 & 17.66{\scriptsize $\pm$0.07} & 16.89{\scriptsize $\pm$0.07} & 16.33{\scriptsize $\pm$0.09} \\
52781.5 & May 21 23:10--23:53 & 17.96{\scriptsize $\pm$0.11} & 17.10{\scriptsize $\pm$0.11} & 16.50{\scriptsize $\pm$0.13} \\
  \hline
%  \end{tabular*}
  \end{tabular}
\end{center}
\end{table*}

The infrared magnitudes of XTE J1720$-$318 decayed less 
during our observation period than the X-ray flux (figure 2).  
During the first 55 days from 2003 January 18, 
the $J$, $H$, and $K_{\mathrm s}$ brightness dropped by $\sim$1 mag (table 1), 
whereas the X-ray flux decreased by a factor of 8 
during the same period.  
This can be explained as being a consequence of the accretion-disk 
instability model, 
in which the optical--IR flux is due to an X-ray irradiated disk 
whose outer part can be kept hot enough to emit the optical flux long 
until the cooling front from inside reaches the outer part 
(Lasota 2001). 

The infrared flux decayed steadily, in general, 
but showed a small increase on January 29 (20 days after January 9), 
and a larger increase around February 17 (39 days after January 9).  
These two flux increases are also observed 
in the X-rays, 
and are probably classified as secondary maxima, 
such as glitches and bumps according to Chen et al. (1997).  
However, such two secondary maxima are not usually recorded in 
XNe on a short time scale like this, 
although a ``tertiary'' maximum was observed for A 0620$-$00, 
only in the optical wavelengths \citep{kuul98}.  
We have not found a good explanation for these multiple flux increases 
in the X-ray and infrared regions.  
The first infrared flux increase (``bump'') 
occurred $\sim$5 days earlier than the X-ray flux, 
although the large errors and poor sampling in the X-ray measurements 
at these times make the comparison difficult.  
Such a time lag is often interpreted as being an ``outside-in''
type outburst.  
The second infrared flux increase (``glitch''), on the other hand, 
seems to have 
occurred at the same time as the X-ray increase, 
and the first and second flux increases might have been caused 
by different mechanisms. 

The observed $J-H$ and $H-K_{\mathrm s}$ colors did not change very much 
over the whole observation period, 
and can be regarded as being constant, 0.81 and 0.52, respectively.  
We noted no significant change in these colors, 
although possibly ($<2\sigma$) bluer $H-K_{\mathrm s}$ colors 
(i.e., smaller $K_{\mathrm s}$ flux increase) were recorded 
when these secondary flux increases occurred.  

The general decay of the infrared light curve can be classified 
as a possible FRED 
(``possible'' because there were no rise phase data; 
Chen et al. 1997), 
which is the most common type of optical light curve.  
The $e$-folding time that was determined from the data 
before the large flux increase before February 17 is $\sim$60 d, 
and it was comparable to the average of the optical $e$-folding time 
for XNe, 67.6 d (Chen et al. 1997).  
We should note, however, that the last observation point on May 21 
shows that the infrared flux decay slowed down (figure 2), 
but, again, no significant change in $J-H$ and $H-K_{\mathrm s}$ compared with the 
previous colors was observed on May 21.  

Let us estimate the extinction that the infrared flux of XTE J1720$-$318 
suffers.  
According to a summary by van Paradijs and McClintock (1995), 
the optical spectra of XNe are indistinguishable from 
spectra of persistent low-mass X-ray binaries; 
they are the spectra of X-ray irradiated accretion disks, 
consisting of blue continua ($T_{\mathrm{eff}} \sim$ 25000--30000 K) 
and a few emission lines.  
Therefore, although the intrinsic infrared colors of XNe 
have not been observationally well determined yet, 
we assume here that the $(J-H)_0$ of XTE J1720$-$318 is $\sim$0 
(high temperature blackbody, neglecting the stellar contribution).  
Then, $E(J-H)$ is $\sim$0.81, and $A_V$ is $\sim$7.8 
if we assume the van de Hulst no. 15 reddening curve \citep{glas99}.  
Similarly, assuming $(H-K_{\mathrm s})_0 \sim$0, 
$E(H-K_{\mathrm s})$ is $\sim$0.52, and $A_V$ is $\sim$9.2 
if we use the $K_{\mathrm s}$ to $K$ extinction ratio in Dutra et al. (2002).  
This is slightly larger, 
and might mean some contribution from other emission than the 
irradiated disk to the $K_{\mathrm s}$ band (see below).  

Unfiltered CCD images taken by one of us (BM) on 2003 January 16.10 
did not detect any 
object brighter than $R = 18.0$ or $I = 16.5$ at the position of the 
infrared counterpart.  This invisibility is consistent with the 
extinction ($A_R \sim$ 6.1 and $A_I \sim$ 4.6) derived from $E(J-H)$. 

In this direction, the $E(B-V)$ estimate based on 
the 100-$\mu$m dust emission by Schlegel et al. (1998) 
is 2.24, and so $A_V$ is $\sim$7.0 mag.  
The relatively high Galactic latitude of 
this line of sight 
(XTE J1720$-$318 is at $l =$\timeform{354D.6}, $b =$\timeform{3D.1}) 
favors a scenario whereby extinction by a Galactic dust layer occurs 
mainly in our neighborhood.  
A simple model calculation (e.g., such as Wainscoat et al. 1992) 
with a dust distribution height scale of 100 pc tells us 
that 50, 75, and 90 \% of the Galactic extinction occurs 
within 2.5, 4.5, and 6.8 kpc from us, respectively.  
Therefore, if the source is more than several kpc away, 
the $A_V$ estimate above is not inconsistent with the 
map of Schlegel et al. (1998).  
It also agrees with a hydrogen column density of 
$1.3 \times 10^{22} {\rm cm}^{-2}$, estimated from 
the XMM-Newton spectrum (Gonzalez-Riestra et al. 2003).  
In this line of sight, 
extinction is also derived from the 2MASS data (Dutra et al. 2003) 
$A_{K, \mathrm{2MASS}} \sim 0.5$ ($A_V \sim 6$), 
somewhat smaller than the above.  

If we assume $A_V$ = 7.8 and $(V - J)_0 = 0$, 
we then find an X-ray to optical flux ratio of a few hundred 
at the beginning of our observations around January 18, 
consistent with the average ratio of $\sim$500 found for low-mass 
X-ray binaries by van Paradijs and McClintock (1995).  

Although some XNe indicate a significant contribution from 
the flat synchrotron emission to the near-infrared fluxes 
(e.g., 
\cite{fend01}; XTE J1859+226: \cite{broc02}; GX339$-$4: \cite{corb02}), 
an irradiated thermal disk seems to dominate in the case of 
XTE J1720$-$318 
because the radio flux detected by \citet{rupe03} was not large. 
Since the radio observation was made 
when the source was in the high/soft state, 
the weak radio synchrotron emission was not surprising.  
However, the detected $K_{\mathrm s}$ band flux is slightly larger ($\sim$0.1 mag) 
than the hot irradiated disk emission only 
[for which $(J-K_{\mathrm s})_0 \sim$0 and $(H-K_{\mathrm s})_0 \sim$0], 
if we assume $A_V$ $\sim$7.8 based on the $J-H$ color; 
this might be due to the synchrotron emission contribution.  

The peak X-ray luminosity of XNe spans from 
$10^{36}$ to $10^{40}$ erg~s$^{-1}$ \citep{chen97}, 
but the typical peak luminosity of the high/soft state of 
the black hole XNe (FRED-type) seems to be $10^{38}$ erg~s$^{-1}$.  
Assuming this value, 
the observed flux 400 mCrab indicates that XTE J1720$-$318 lies at 
$d\sim$10 kpc, which puts it among the most distant XNe 
detected, such as GRS 1915+105 at 12.5 kpc and 4U 1730$-$335 at 10 kpc 
\citep{tana96}.  

In the direction toward the Galactic center, 
the optical observations were limited to relatively nearby 
XNe.  
Our observations, made over 130 days in the $J, H,$ and $K_{\mathrm s}$ bands 
for a typical XN possibly beyond the Galactic center, 
demonstrate that near-infrared observations provide us 
with data to be compared with the X-ray light curve, 
even for such XNe which suffer heavy optical extinction.  

\bigskip
We would like to express our thanks to the staff 
of South African Astronomical Observatory for their 
kind support during the observations.  
This work is partly supported by Grant-in-Aid 
for Scientific Research of the Ministry of Education, 
Culture, Sports, Science, and Technology.

\end{document}